\documentclass{aa}

\usepackage{graphicx}
\usepackage{caption}

\usepackage{txfonts}

	\usepackage[utf8]{inputenc}
	\usepackage{amsmath}
 \usepackage{amssymb}
\usepackage[colorlinks=true,linkcolor=bluea,allcolors=blue]{hyperref}

\begin{document}

   \title{Gas flows in the central region of the Seyfert galaxy NGC\,4593 with MUSE}

   \author{D. Mulumba\inst{1}
          \and J. H. Knapen\inst{2,3}
          \and S. Comerón\inst{3,2}
          \and C. Ramos Almeida\inst{2,3}
          \and W. O. Obonyo\inst{5,1}
          \and M. Pereira-Santaella\inst{4} 
          \and P. Baki\inst{1}
          }

   \institute{Department of Physics, Technical University of Kenya, Nairobi, Kenya;\\ \email nthokidorcus@gmail.com
   \and Instituto de Astrofísica de Canarias E-38205, La Laguna, Tenerife, Spain;
   \and Departamento de Astrofísica, Universidad de La Laguna E-38206, Tenerife, Spain;
   \and Instituto de Física Fundamental, CSIC, Calle Serrano 123, 28006 Madrid, Spain;
   \and Department of Mathematical Sciences, University of South Africa, Cnr Christian de Wet Rd and Pioneer Avenue, Florida Park, 1709, Roodepoort, South Africa.}
   \date{}

  \abstract 
    {Understanding how gas flows into galactic centres, fuels the active galactic nucleus (AGN), and is in turn expelled back through feedback processes is of great importance to appreciate the role AGN play in the growth and evolution of galaxies.}
   {We use Multi Unit Spectroscopic Explorer-adaptive optics (MUSE-AO) optical spectra of the inner $7\farcs5 \times 7\farcs5$ ($\mathrm{1.3\,kpc\times 1.3\,kpc}$) of the nearby Seyfert 1 galaxy NGC\,4593 to characterise its ionised gas kinematics.}
   {We fitted single-Gaussian components to the [O\,\textsc{iii}]\,$\lambda$5007 and [N\,\textsc{ii}]\,$\lambda$6583 emission lines, and double-Gaussian components to H$\alpha$ and H$\beta$ to determine the main ionisation mechanism of the gas. To determine the kinematics of the ionised gas, we fit double-Gaussian components to the [O\,\textsc{iii}]\,$\lambda$5007 line.}
{The high angular resolution MUSE data ($\sim0\farcs12=20\,{\rm pc}$) capture structures of the circumnuclear region including the innermost spiral that feeds the nucleus. Based on the stellar kinematic maps, we confirm the presence of a rotating disc, whilst for the ionised gas, we find high-velocity dispersion values of up to $\mathrm{200-250\,km\,s^{-1}}$ that show that part of the gas is highly perturbed. The dominant ionisation mechanism of the gas is AGN photoionisation, which reaches the highest values within the innermost 4\arcsec (680\,pc) diameter of the galaxy. At larger radii, the emission line ratios correspond to values in the composite region of the Baldwin, Phillips and Terlevich (BPT) diagram.}
{The broad-component of [O\,\textsc{iii}]\,$\lambda$5007 shows blue-shifted velocities on the east side of the central 2\arcsec (340\,pc), which spatially coincide with a region of high velocity-dispersion. This confirms the presence of outflowing gas. We estimate a mass outflow rate and kinetic power of ${\dot{M}}\geq {0.048 \,M_\odot\,{{\rm yr}^{-1}} }$ and ${\dot{E}_{\text{kin}}} \geq 4.09 \times 10^{39} \, \text{erg} \, \text{s}^{-1}$. The derived mass outflow rate is consistent with that expected from empirical relations between mass outflow rate and AGN luminosity for a low-luminosity AGN such as NGC~4593. High angular resolution integral field observations can enable multi-component analysis of the innermost regions of galaxies, allowing a detailed view of ionised gas flows.}   
   \keywords{Galaxies: active -- Galaxies: Individual: NGC\, 4593 -- Galaxies: Kinematics and dynamics -- Galaxies: nuclei }
\authorrunning{D. Mulumba et al.}
  \maketitle
  %
\section{Introduction} 
 Gas flows into the centres of galaxies are an important ingredient in the formation, growth and evolution of galaxies \citep{knapen2019muse,storchi2019observational}. The process of gas falling into the central supermassive black hole (SMBH) can lead to high levels of non-stellar emission which results in an AGN. However, the details of how these gas flows reach the centre of a galaxy and how they in turn affect  their surroundings through feedback is not well understood, despite ongoing research  \citep{2024Galax..12...17H}.

Violent disturbances caused by major mergers can drive gas to the centre of galaxies and fuel starbursts (see \citealp{storchi2019observational}). Shocks and gravitational torques associated with bars also lead to inward gas transportation  (\citealp{simkin1980nearby}; \citealp{sakamoto1999bar}). However, in the presence of inner Lindblad resonances (ILRs), the gas does not flow further towards the galactic nucleus, but accumulates in rings near the ILRs because of positive gravitational torques (\citealp{knapen1995central}; \citealp{piner1995nuclear}).   

Once inside the inner kiloparsec, further inflow inwards on the scales of hundreds and tens of parsec is influenced by smaller non-axisymmetries (\citealp{begelman1984theory}; \citealp{shlosman1989bars}). The accumulation of gas in the central parts of the galaxy may lead to both starburst and AGN formation, and if the two occur simultaneously, such a galaxy is called ‘composite' (see \citealp{knapen2019muse}).
   
The supply of gaseous fuel feeding the AGN may regulate nuclear activity. There is a tight correlation between the SMBH mass and the main properties of its host galaxy, including its  luminosity, mass and the velocity dispersion of the bulge (\citealp{ferrarese2000fundamental}). This shows that there is co-evolution between the SMBH and its host galaxy (\citealp{kormendy2013coevolution}; \citealp{heckman2014coevolution}). Feedback from the AGN in the form of jets, radiation, and accretion disc winds 
may regulate the growth  of a galaxy by suppressing star formation due to gas that is ejected from the nuclear region, is heated, or is just disrupted \citep{2024Galax..12...17H}.
   
In order to understand the transport of gas in to the AGN and also how it affects its surroundings through feedback, we study the stellar kinematics and the emission line properties of the interstellar gas around the nucleus of NGC\,4593. This study requires observations with a very high spatial resolution (e.g.- \citealt{comeron2021complex,bessiere2022,ramos2022}). With the advent of AO-supported integral field spectroscopy (IFS), the detailed study of the central tens of parsecs of many galaxies becomes possible. We used high angular resolution archival IFS data obtained with the AO narrow-field mode (NFM) of the MUSE; \citealp{bacon2010muse}) at the Very Large Telescope (VLT).

NGC~4593 is a nearby spiral galaxy that is also known as Mrk~1330. It is an active galaxy classified as Seyfert~1.3 of morphological type (R$^{\prime}$)SB(rs)\,b at a distance of $35.4\pm3.0$\,Mpc, with a diameter 39.3\,kpc and an angular size of $2\farcm75  \times 2\farcm29$ (\citealp{de1991third}). At this distance, one arcsecond corresponds to 170\,pc. NGC~4593 hosts a SMBH at its centre, with a mass of $M_{\rm BH} = ( 9.8 \pm 2.1) \times 10^6\,{M_\odot}$ (\citealp{bentz2006reverberation}). Variability in the X-ray band emission was observed (\citealp{burnell1979ngc}) and far-infrared observations revealed a halo-like morphology that is slightly elongated along a position angle of $100^{\circ}$ \citep{mchardy2018x}. \citet{delgado1997h} noted the presence of dust and a blue nuclear ring with a radius of about 2$^{\prime\prime}$. Observations with Hubble Space Telescope/ Near Infrared Camera and Multi-Object Spectrometer 2(HST/NICMOS2) confirm that the central region is dominated by spiral dust lanes surrounding a pseudo-ring without evidence of an inner bar (\citealp{marin2007atlas}). Observations from \cite{ruschel2021agnifs} showed extended [O\,\textsc{iii}]\,$\lambda$5007 emission whose peak is displaced by about $0\farcs3$ (40\,pc) to the north-eastern side of the nucleus. They also reported the tentative detection of an ionised outflow based on the higher velocity dispersion values of the [O\,\textsc{iii}]\,$\lambda$5007 emission ($W_{80}\sim400-500\,{\rm km\,s}^{-1}$) and gas densities ($n_{\rm e}\sim200-600\,{\rm cm}^{-3}$) that they measured in the east to south-eastern direction. 

\section{Observations and data processing}

\subsection{MUSE-AO observations}

We used archival data from the MUSE IFS at the VLT UT4. MUSE contains 24 integral field units (IFUs) and uses the image-slicing technology to deliver very high quality spectroscopic data (\citealp{bacon2010muse}). It is enhanced with a deformable secondary mirror (DSM) that implements all the VLT-required functionalities and has ground-layer AO wavefront sensing modules such as the ground atmospheric layer adaptive corrector for spectroscopic imaging (GALACSI). Attached to the main structure of the telescope is the four laser guide-star facility (4LGSF), which provides the hardware and software to launch the laser beams into the atmosphere (\citealp{stuik2006galacsi}). In MUSE, the NFM has a wide wavelength range of $4750\,\AA - 9350\,\AA$ and a field of view (FOV) of $7\farcs5 \times 7\farcs5$ with a sampling of 0.0253\,arcseconds per pixel. The angular resolution, obtained from measuring the full width at half maximum (FWHM) of stars in the FOV, is $\sim$0\farcs12 (20 pc). The data used in this work were taken on August 24, 2019, within the programme ID.0103.B – 0908 (Principal Investigator: K. Jahnke). The total exposure time was 2400\,s in a single observation block. We used a fully reduced (\citealp{delmotte2016validation}) science-ready data cube provided by  European Southern Observatory's (ESO) Phase 3 data products. This data cube has two extensions, one containing the signal and the other containing the error estimates. The fully reduced three-dimensional data cube has two spatial and one wavelength axes.

\subsection{MUSE data processing}
\label{processing}

\begin{figure*}[!htb]
    \minipage{0.30\textwidth}
    \includegraphics[width =\linewidth]{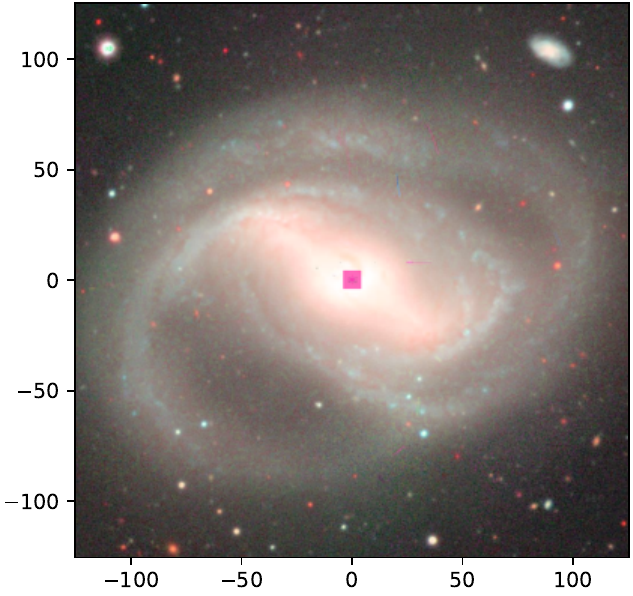}
      \endminipage
    \hfill
    \minipage{0.30\textwidth}
    \includegraphics[width = \linewidth]{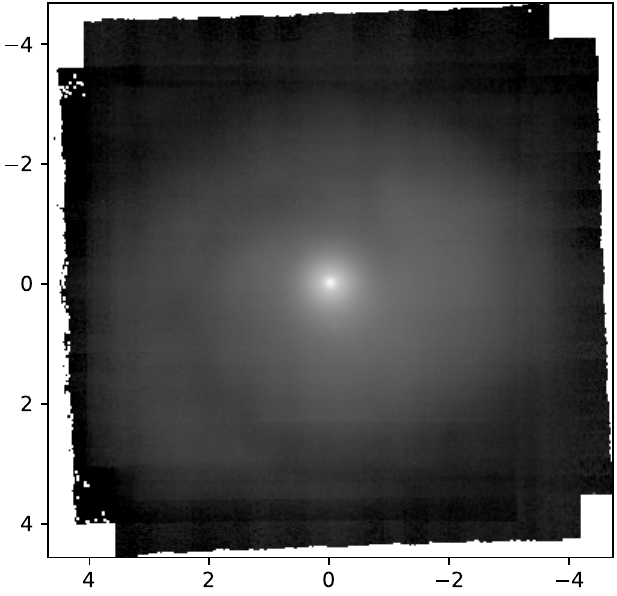}
         \endminipage
       \hfill
    \minipage{0.28\textwidth}
    \includegraphics[width = \linewidth]{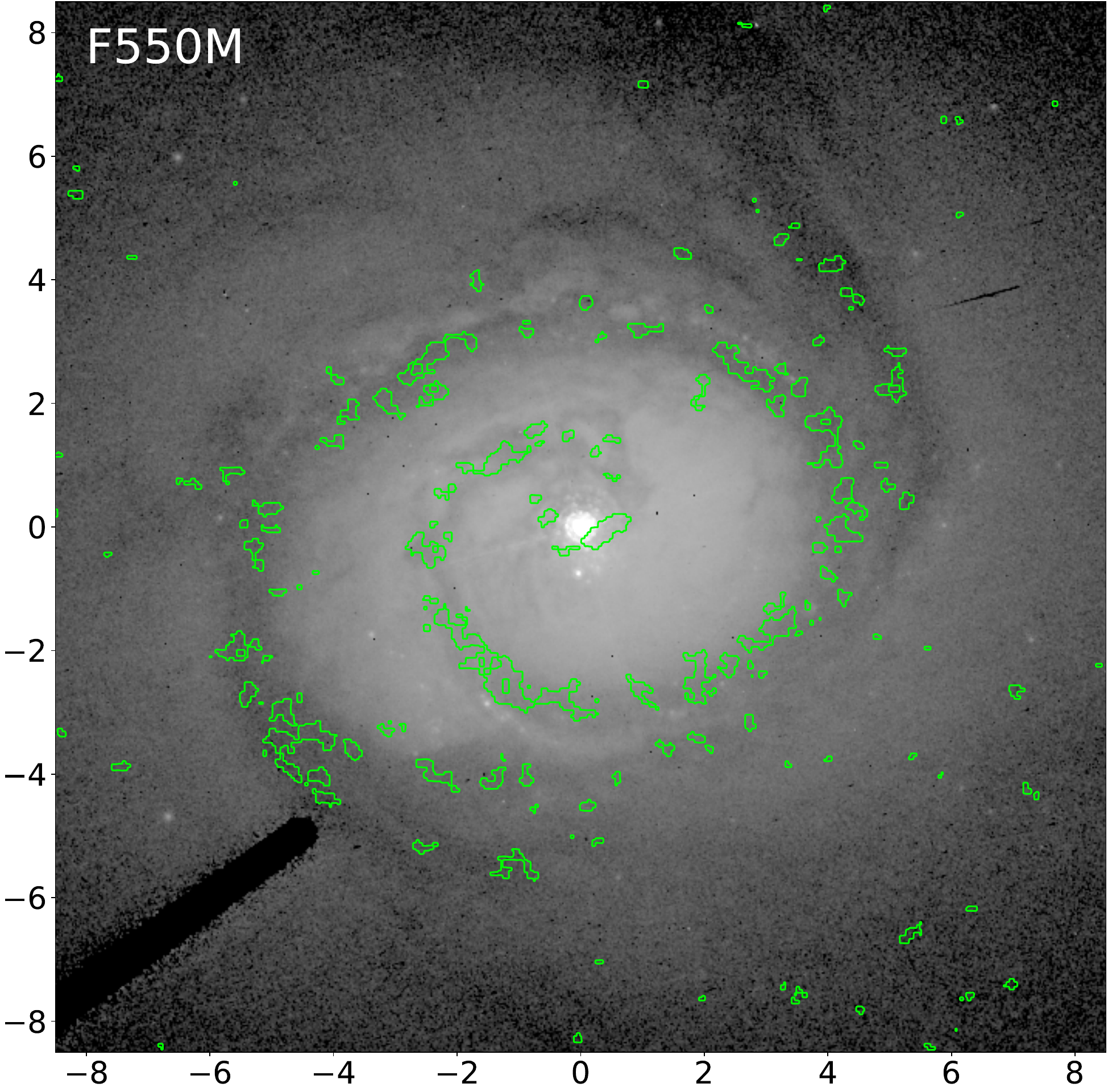}
         \endminipage   
    \caption{NGC 4593. Left panel: Archival Dark Energy Spectroscopic Instrument (DESI) legacy survey image of the galaxy (\citealp{dey2019overview}). The purple square at the centre shows the FOV of the MUSE-NFM observations ($7\farcs5 \times 7\farcs5$). Middle panel: MUSE image of the galaxy collapsed in the spectral range 5300--5750 $\AA$.  Right panel: HST/ACS image of the galaxy in the F550M filter. The green contours correspond to the CO(1-0) line emission from ALMA. The molecular gas distribution spatially coincides with the spiral dust lanes. All images show a bright nucleus and a single spiral arm extending right into the centre of the galaxy. North is up, east to the left and the scale of each panel is in arcseconds.}
    \label{Fig.1}
\end{figure*}
We used the galaxy IFU spectroscopic tool (\texttt{GIST}), version 3.1.0 to process the data cube. It is a comprehensive 
Python-based framework for the scientific analysis of fully reduced IFU spectroscopic data (\citealp{bittner2019gist}). The input data cube was read using \texttt{GIST} and the wavelength range is shortened to $4750\,\r{A}- 7000\,\r{A}$. This region contains the emission lines of interest, for instance to produce the BPT diagrams to we used to study the ionisation mechanisms of the central region of the galaxy. Above $7000\,\r{A}$, there is contamination due to strong sky lines, and the range $5780\,\r{A} - 6050\,\r{A}$, which is affected by the laser guide stars was,ignored. Spatial masking was performed to remove defunct spaxels and those with a signal-to-noise ratio (S/N) below the given threshold before we binned the data.
Our data contain both high and low S/N associated with line emission and stellar absorption respectively, we therefore used two binnings produced with the Voronoi binning code written by \cite{cappellari2003adaptive}. For the stellar  component we computed the median S/N from the rest-frame wavelength range 5580 -- 5680\,\AA, which does not include any strong emission lines. We required an ${\rm S/N}=50$ per stellar bin and a total of 426 Voronoi bins were generated. For the emission line binning, we  required an ${\rm S/N}=100$ and in this case, a total of 762 Voronoi bins were generated. The emission line or ionised gas binning was made based on a [O\,\textsc{iii}]\,$\lambda$5007 continuum-subtracted image. We built the image by integrating the data cube flux in a $10\,\AA$  window centred on the [O\,\textsc{iii}]\,$\lambda$5007 rest-frame wavelength of NGC~4593, and by subtracting the integrated flux in a window of equal width but $50\,\AA$ redwards. We considered spaxels with an S/N greater than 0.1 and 1.0 for the emission line and stellar binning, respectively. Inevitably, some isolated spaxels might have an S/N above the threshold due to random fluctuations. To avoid these spaxels from being accreted into Voronoi bins we modified GIST to consider clusters with a minimum of 50 contiguous spaxels for the binning. The number of 50 spaxels is arbitrary.

We used the penalized pixel-fitting (\texttt{pPXF}) method to extract the stellar velocity field and velocity dispersion. The code performs a full spectral fitting of each stellar emission bin with a linear combination of template spectral energy distributions widened to match the line-of-sight velocity distribution (LOSVD) of the spectral lines. We used the E-MILES library (\citealp{vazdekis2016uv}) templates together with BaSTI isochrones (\citealp{pietrinferni2004large}), a Kroupa universal stellar initial mass function (\citealp{kroupa2001variation}), and solar abundances. The initial guess for the velocity dispersion was $\mathrm{50\,km~s^{-1}}$ and the stellar continuum was modelled with an eighth-degree additive Legendre polynomial. Very strong sky and emission lines were masked with windows of various widths: [N\,\textsc{i}]\,$\lambda$$\lambda$ 5198, 5200,  [He\,\textsc{i}]\,$\lambda$5875, [O\,\textsc{i}]\,$\lambda$$\lambda$6300,6363, [N\,\textsc{ii}]\,$\lambda$$\lambda$6547,6583, and [S\,\textsc{ii}]\,$\lambda$$\lambda$6716,6730 were masked with a 30\,\AA\, width; H$\beta$ with 75\,\AA; [O\,\textsc{iii}]\,$\lambda$4958 with 20\,\AA; [O\,\textsc{iii}]\,$\lambda$5007 with 40\,\AA; H$\alpha$ with 130\,\AA; and the sky line at $\lambda$5577 with 20\,\AA. We also excluded a 280\,{\AA} window centred on  5914\,{\AA} to avoid being affected by the laser guide stars.

To study the ionised gas, we used the Gas and Absorption Line Fitting\texttt{PyGandALF} module \citep{sarzi2006sauron,falcon2006sauron} from the \texttt{GIST} pipeline, which was run on the bin level to perform the emission line analysis. Since NGC\,4593 is a Seyfert 1.3 galaxy showing broad-components associated with the broad line region (BLR) in the permitted lines (e.g. H$\alpha$ and H$\beta$), a narrow-component from the narrow-line region (NLR), and a broad-component associated with the ionised gas outflow \citep{ruschel2021agnifs}, we focused on the [O\,\textsc{iii}]\,$\lambda$5007 line to study the properties of the outflowing gas, since it is a strong forbidden transition that is not in the vicinity of other strong emission lines with a BLR component, such as H$\alpha$. Hence, any broadening of the [O\,\textsc{iii}] line profiles will be associated with kinematically disrupted gas, and not with the BLR. Therefore we only performed a two-Gaussian component fitting for the [O\,\textsc{iii}] line when we tried to characterise the outflow kinematics (see Section \ref{emission}). For [N\,\textsc{ii}] were unable to do this because they are blended with H$\alpha$ and fully immersed in the broad-component from the BLR. We therefore just a single Gaussian-component to fit the [O\,\textsc{iii}] and [N\,\textsc{ii}] lines in Section \ref{bpt}. The initial guess for the velocity and velocity dispersion is 0 and $\mathrm{100\,km~s^{-1}}$ for the narrow-component, while for the broad outflow and BLR components, the initial guess is -500 and $ \mathrm{500\,km~s^{-1}}$ respectively.

We reduced the number of degrees of freedom in the fits by fixing the flux ratios of several doublets to their theoretically estimated values 
(\citealp{storey2000theoretical}), which are listed in Table \ref{table 1}. We also tied the kinematics of all the lines considered here to those of [O\,\textsc{iii}], since otherwise it would be impossible to successfully fit the H$\alpha$+[N\,\textsc{ii}] blend. Due to the strong AGN emission, the stellar contribution to the emission lines is minimal compared to that from the emission lines throughout the whole FOV. In any case, we used the stellar fit from \texttt{pPXF} as input to \texttt{PyGandALF}, multiplied by a second-order Legendre polynomial. This component accounts for the underlying stellar and AGN continuum. 
\begin{table}
 \caption{Fixed emission line flux ratios.}
     \centering
     \begin{tabular}{c|c}
     \hline \hline
  Emission lines & Ratio \\ [0.5ex]
       \hline
[O\,\textsc{iii}]\,$\lambda$4959 , [O\,\textsc{iii}]\,$\lambda$5007& 0.335\\
  \(\text{[O\,\textsc{i}]}\)\,$\lambda$6364, [O\,\textsc{i}]\,$\lambda$6300 & 0.330 \\
\(\text{[N\,\textsc{ii}]}\)\,$\lambda$6548, [N\,\textsc{ii}]\,$\lambda$6583 & 0.327\\
     \hline
          \end{tabular}
  \label{table 1}      
\end{table}

\subsection{DESI, {\it HST}, and ALMA data}
The left panel in Fig.~\ref{Fig.1} shows an archival DESI image of the galaxy.
To display the area covered by MUSE-NFM (7\farcs5$\times$7\farcs5), we overlay a square in the centre. With an angular resolution of approximately $0\farcs05$ and a pixel size of $0\farcs025$, the archival HST image in the right panel of Fig. \ref{Fig.1} was captured using the F550M filter of the ACS in the high-resolution channel (HRC) mode \citep{grogin2011candels}.  The green contours in the right panel of Fig. \ref{Fig.1} correspond to the CO(1-0) emission detected with the Atacama Large Millimeter/sub-millimeter Array (ALMA). These data were obtained on November 3, 2021, as part of project 2021.1.00812.S (PI: C. Ricci) with a resolution of $0\farcs189$. Since the astrometry of HST is not accurate, we applied a shift between the HST image and the ALMA CO map so that the dust lanes coincide with the molecular gas, which is normally the case in spiral galaxies \citep{bolatto17,ramos2022}. Khianfar et al. (in prep.) study the CO emission from NGC\,4593 in more detail on the basis of CO(2-1) ALMA data from project 2017.1.00236.S (PI: M. Malkan).

\section{Results}
\label{results}
\subsection{Morphology}
\label{morphology}

In the middle panel of Fig.~\ref{Fig.1}, the MUSE image of NGC 4593 collapsed in the spectral range 5300 – 5750 $\AA$ shows a bright nucleus dominated by AGN emission and an inner spiral arm extending  right into the centre of the galaxy. This coincides with what \cite{eskridge2002near} reported for NGC 4593 in their study of the near-infrared and optical morphology of spiral galaxies. From the similarity of the morphology of the images, it is clear that the spatial resolution of the MUSE data (0\farcs12) is comparable to that of the HST image (0\farcs05). The ALMA CO(1-0) contours shown in the right panel of Fig.~\ref{Fig.1}, which probe cold molecular gas with an angular resolution similar to that of the MUSE data, trace the dusty spiral arms of the galaxy, which were reported by \cite{kormendy2006pseudobulges}.

\subsection{Stellar kinematics}
\label{stellar}

\begin{figure}
    \centering
    \includegraphics[width=0.45\textwidth]{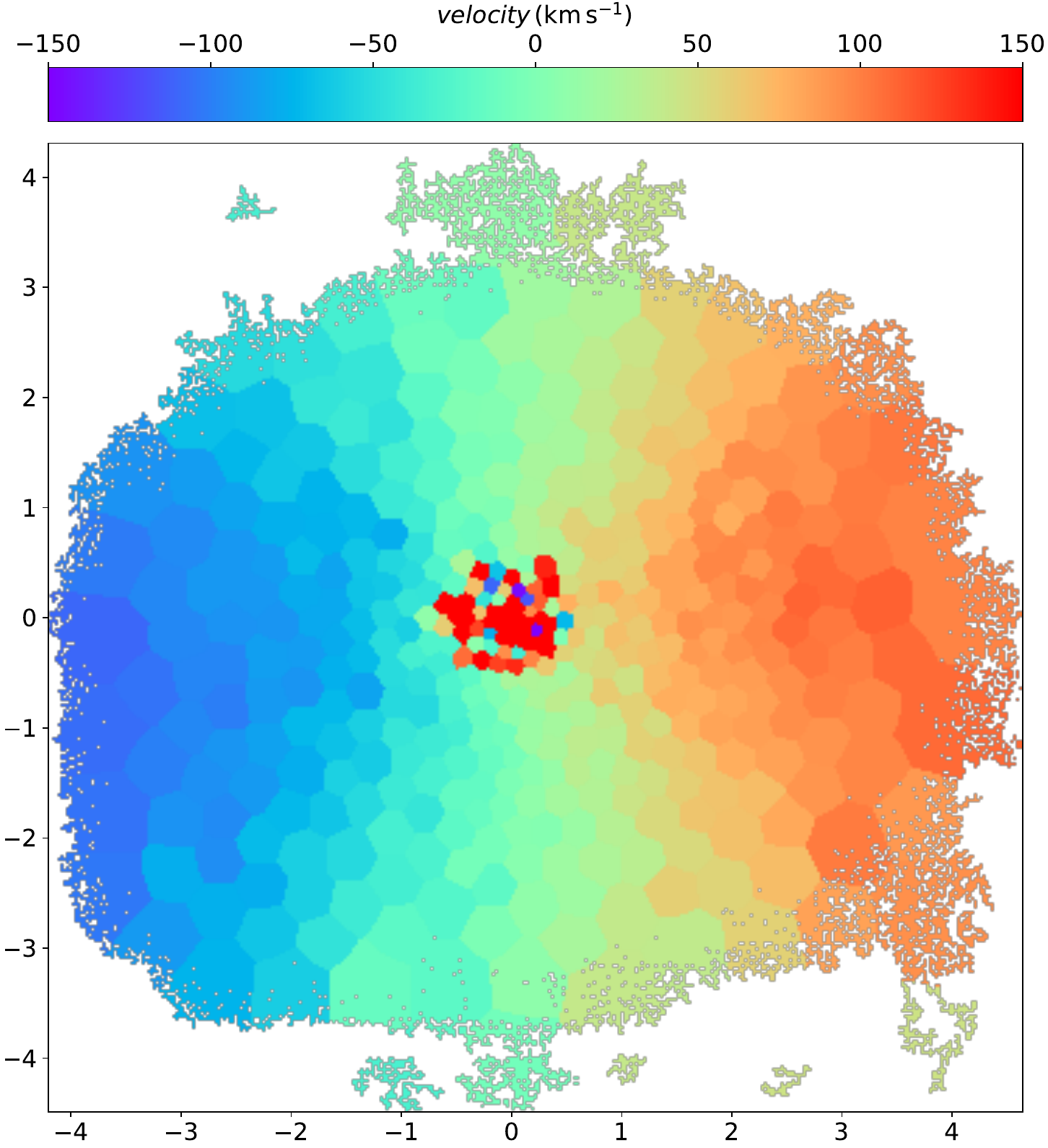}
    \includegraphics[width=0.45\textwidth]{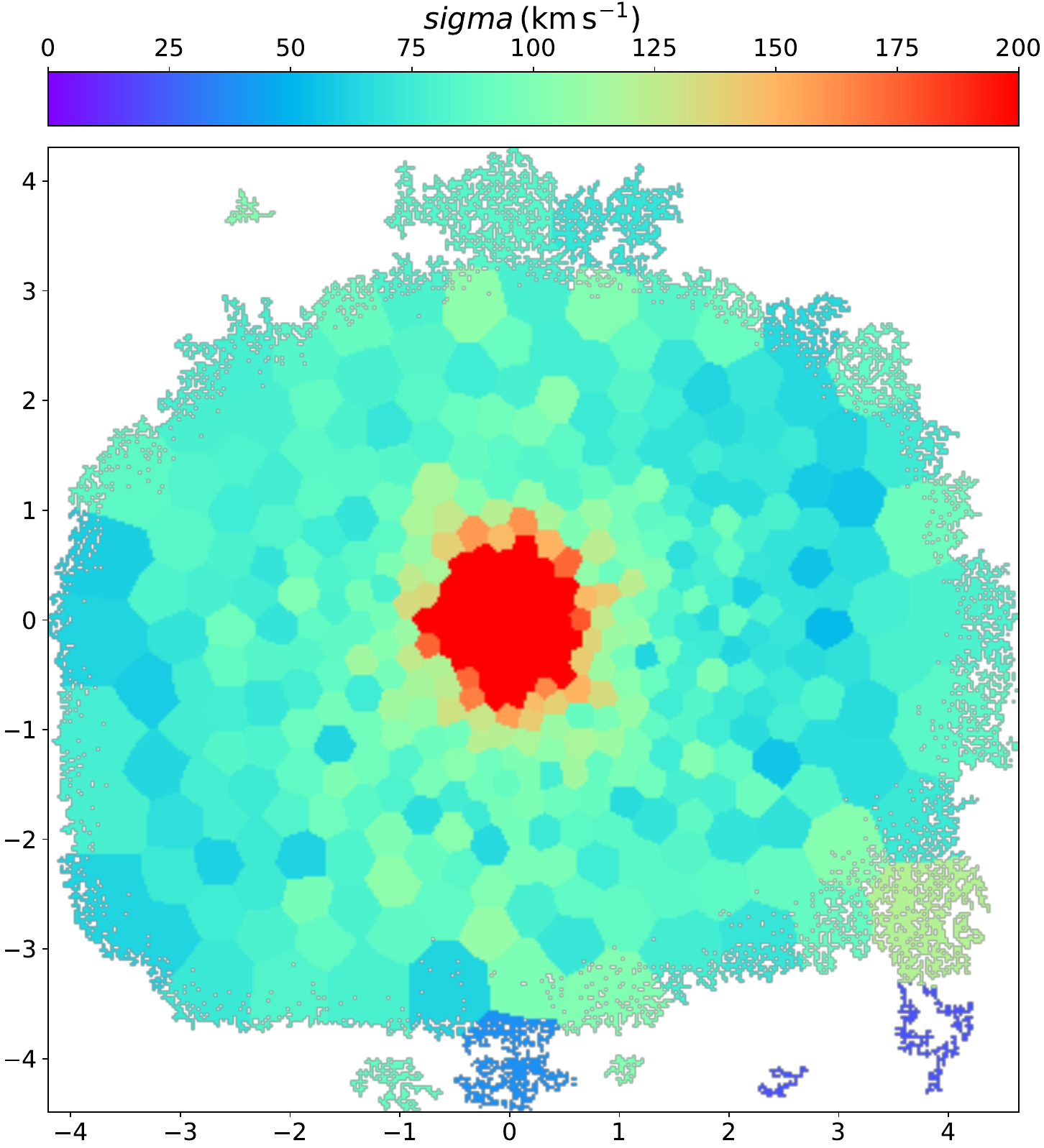}
    \caption{Stellar kinematics maps of NGC~4593. Top panel: Velocity map of the stellar component, where zero velocity corresponds to $z=0.0084$ ($cz = 2518\,{\rm km\,s}^{-1}$). Bottom panel: Velocity dispersion map of the stellar component. The colour scale and units are as shown in the bar above each map. North is up and east is to the left, and the scale of each panel is in arcseconds.}
    \label{Fig.2}

\end{figure}

The stellar velocity and velocity dispersion maps are similar to those of the ionised gas disc (see Sect.~\ref{emission}) and show that the circumnuclear stellar population is supported by rotation with a velocity amplitude of $\pm110\,{\rm km\,s}^{-1}$ (see Fig.~\ref{Fig.2}), which matches the findings by \cite{barbosa2006gemini}. 
It has a kinematic major axis of ${\rm PA}\sim100^{\circ}$, and the east side is blue-shifted and the west side red-shifted, which roughly corresponds to the photometric axis of the inner parts of the galaxy in \citet{kormendy2006pseudobulges}. We were unable to accurately fit the stellar kinematics in the central $\sim 0\farcs7$ (120\,pc) of the galaxy  for any of the maps because the line and continuum emission from the AGN are so strong that the stellar absorption lines are completely diluted. This is shown on the right panel of Fig.~\ref{Fig.2} as a region with unrealistically high velocity dispersion values. For the region outside $\sim 0\farcs7$, we find a typical velocity dispersion of $\sigma=83\pm 4.8\,{\rm km\,s^{-1}}$ which is lower than the value of $\sigma=139\pm5\,{\rm km\,s^{-1}}$ found by \cite{caglar2020llama} for the innermost $1\farcs8$. The reason for the discrepancy is probably a combination of us missing regions with a presumably higher velocity dispersion regions within the inner $\sim 0\farcs7$ and the beam smearing induced by the relatively large aperture in \cite{caglar2020llama}.

\subsection{Emission lines}
\label{emission}
 
As explained in Section \ref{processing}, to study the properties of the ionised gas kinematics, we focused on [O\,\textsc{iii}]\,$\lambda$5007 because any broadening of the line profiles will be associated with kinematically disrupted gas and not with the BLR. Therefore, we performed two Gaussian fits, narrow and broad, which correspond to the narrow and outflow components, respectively.
 
\begin{figure*}
    \minipage{0.32\textwidth}
    \includegraphics[width = \linewidth]{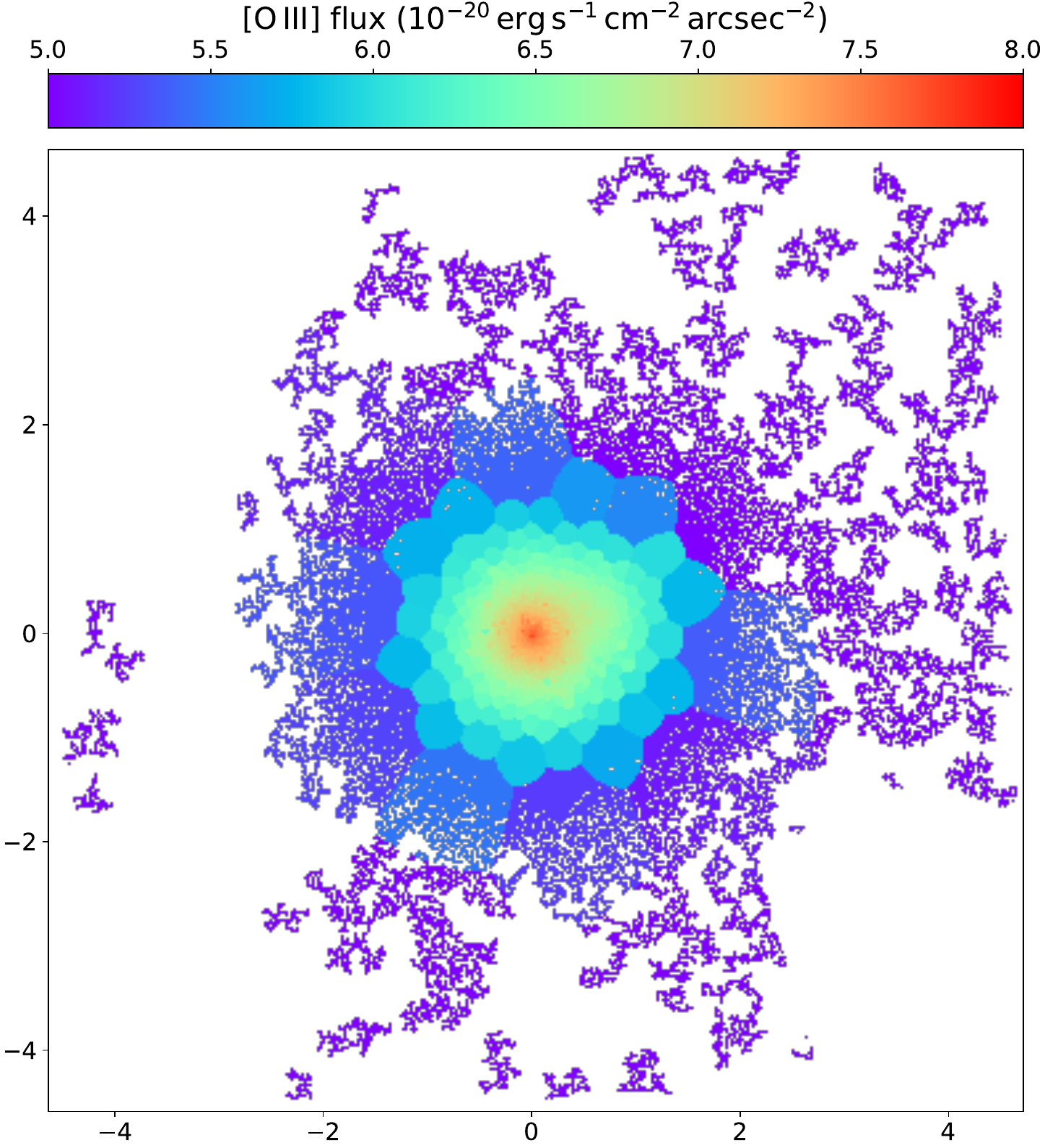}
         \endminipage
    \hfill
    \minipage{0.32\textwidth}
    \includegraphics[width = \linewidth]{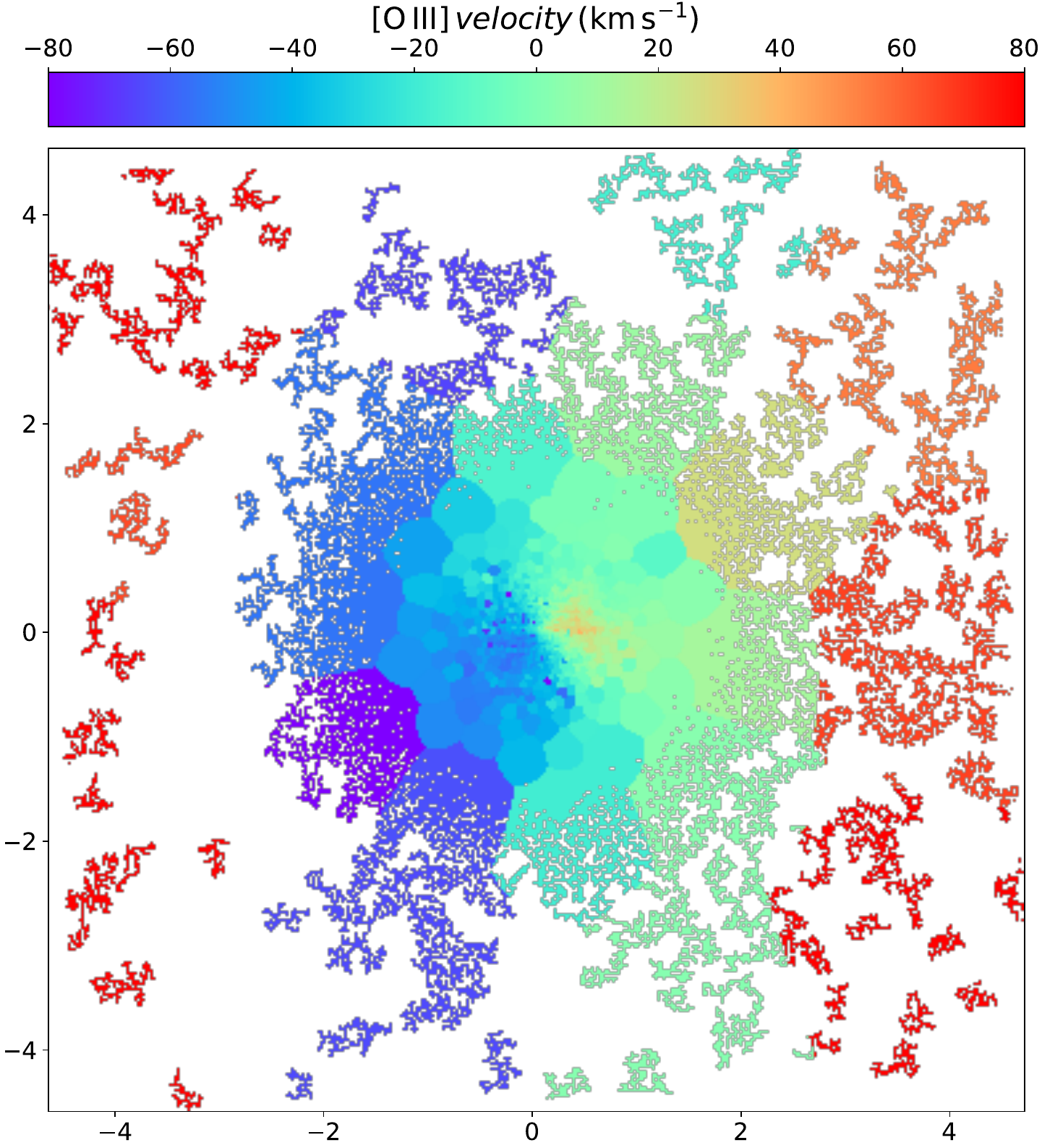}
         \endminipage
     \hfill
    \minipage{0.32\textwidth}
    \includegraphics[width = \linewidth]{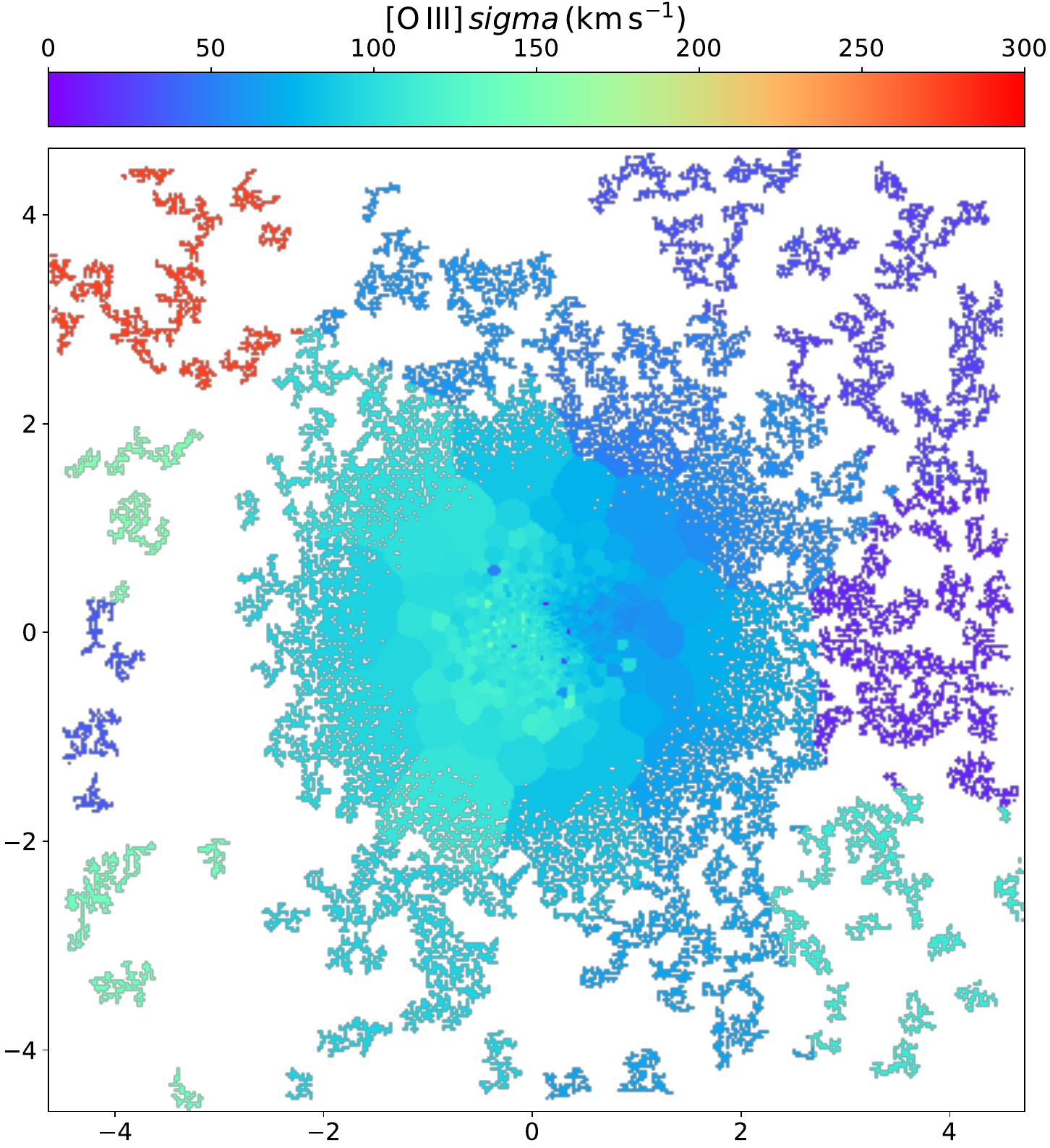}
          \endminipage
  \caption{From left to right: Flux, velocity, and velocity dispersion maps of the [O\,\textsc{iii}]\,$\lambda$5007 narrow-component emission. The colour scale and units are as shown in the colour bar above each map and are the same as in Fig.~\ref{Fig.4} to facilitate comparison. North is up and east is to the left, and the scale for each panel is in arcseconds.}

   \label{Fig.3} 
\end{figure*}

Fig.~\ref{Fig.3} shows the [O\,\textsc{iii}]\,$\lambda$5007 gas flux and kinematics maps of the narrow-component, which traces the ionised gas disc. The flux map shows a strong nuclear peak at the position of the AGN. The flux gradually decreases with increasing radius. 
In the velocity map shown in the middle panel of Fig. \ref{Fig.3} positive velocities are seen to the west and negative velocities to the east. It has a major-axis PA similar to that of the stellar velocity map {\bf (${\rm PA}\approx100^{\circ}$),} although with a smaller velocity amplitude of $\sim90$\,km\,s$^{-1}$.

\begin{figure*}
    \begin{minipage}{0.32\textwidth}
    \includegraphics[width = \linewidth]{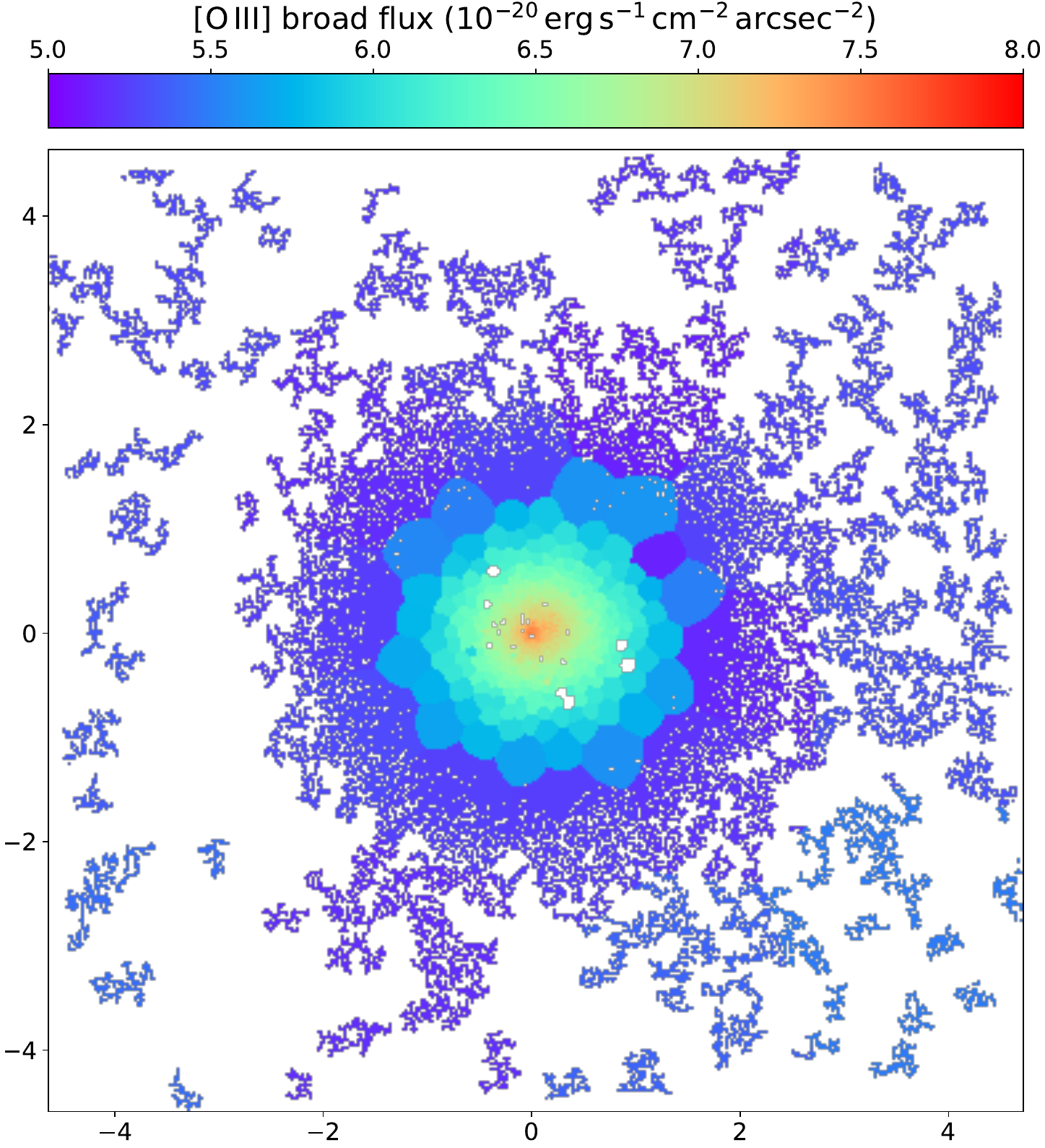}
          \end{minipage}
    \hfill
    \begin{minipage}{0.32\textwidth}
    \includegraphics[width = \linewidth]{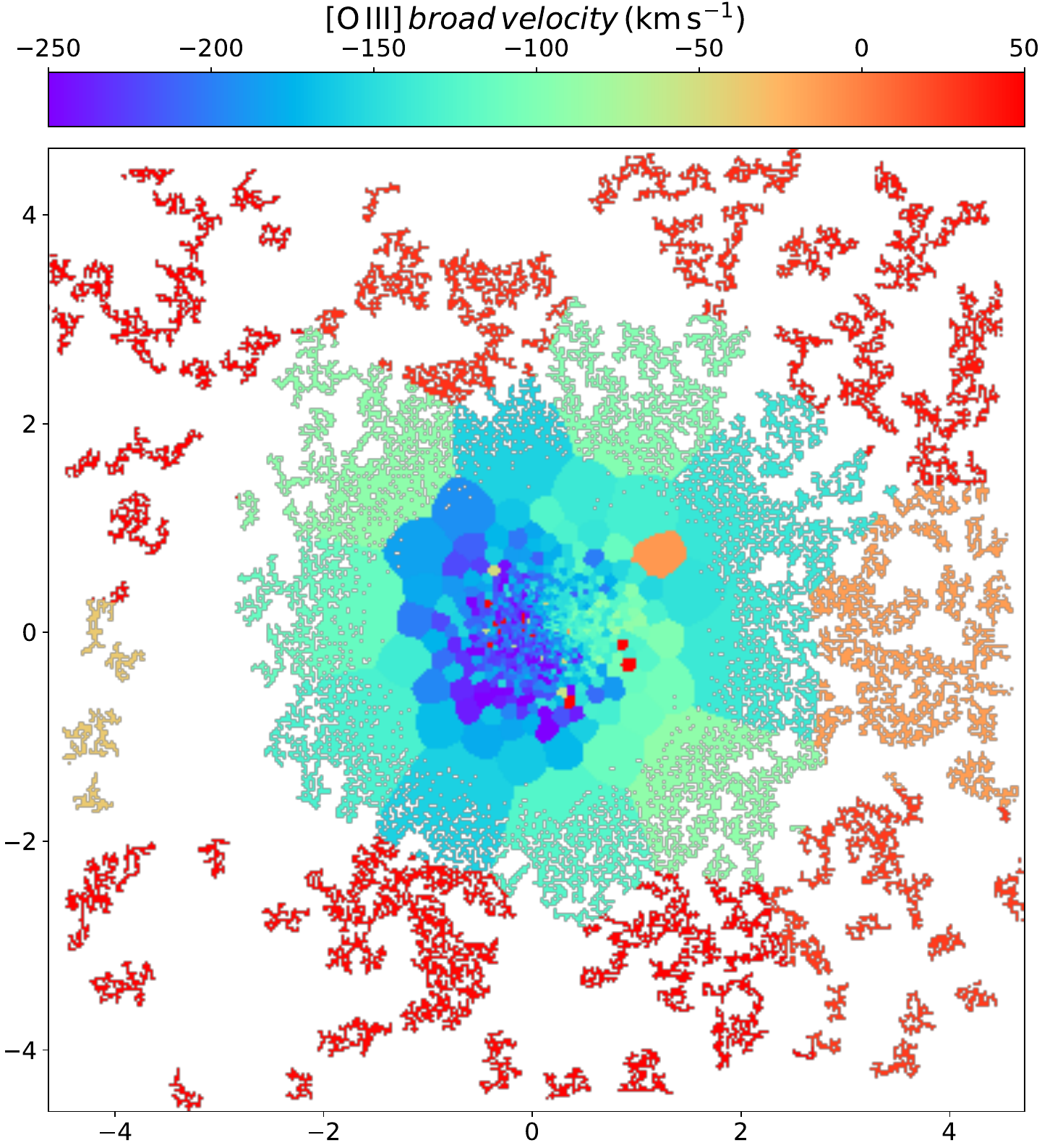}
         \end{minipage}
     \hfill
    \begin{minipage}{0.32\textwidth}
    \includegraphics[width = \linewidth]{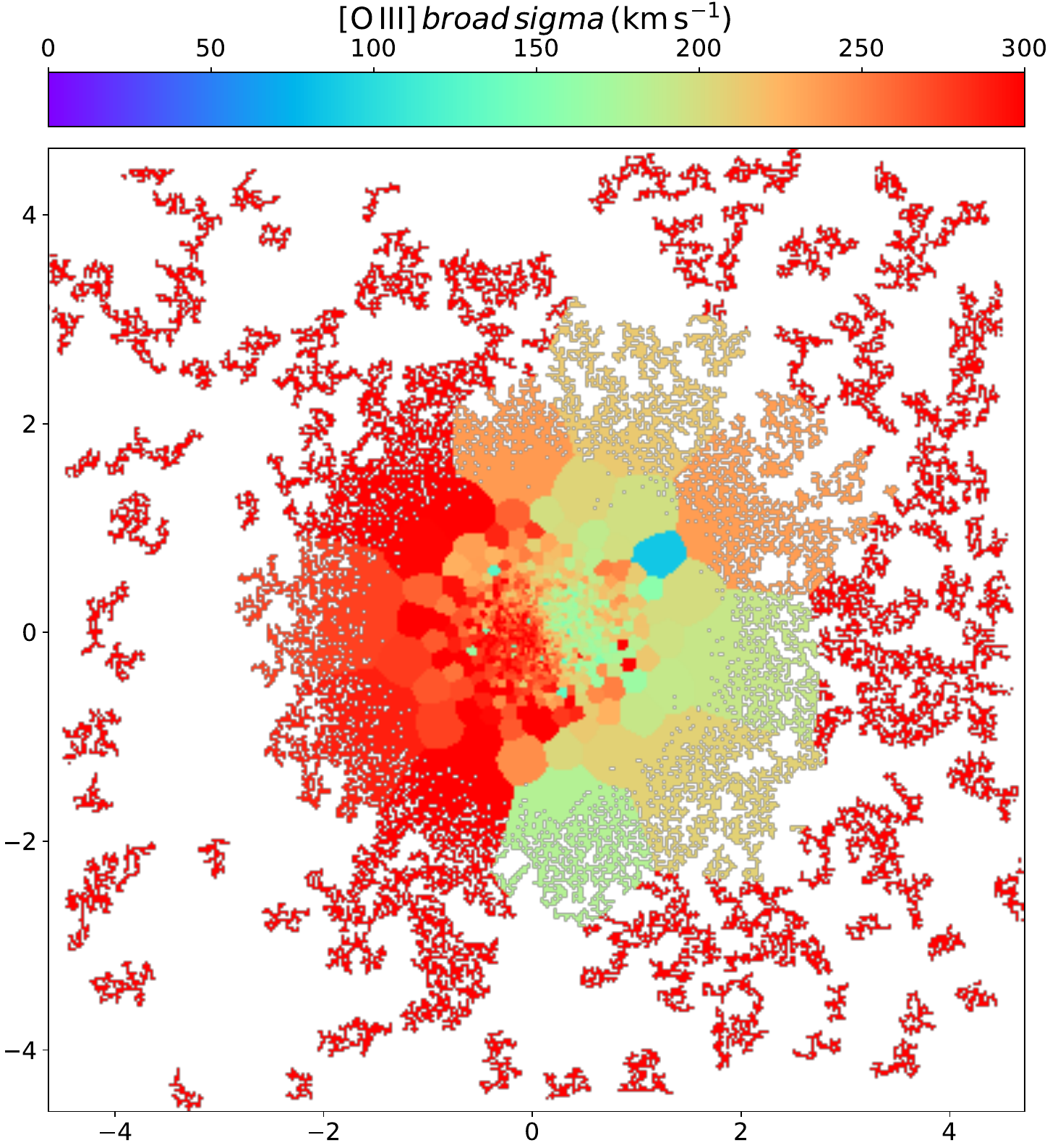}
         \end{minipage}
         
 \caption{As Fig.~\ref{Fig.3}, now for the broad-component emission.}
     \label{Fig.4}
\end{figure*}

Fig.~\ref{Fig.4} shows the [O\,\textsc{iii}]\,$\lambda$5007 gas flux and kinematics maps for the broad-component, which corresponds to kinetically disrupted gas. The flux maps also show a strong nuclear peak, the intensity of which decreases with increasing radius. The kinematic maps show blue-shifted velocities of up to $-200$\,km\,s$^{-1}$ on the east side of the galaxy, within a radius of $\sim$2\arcsec (340\,pc). This coincides with a region of high velocity dispersion (up to {\rm 250}\,km\,s$^{-1}$), confirming the presence of outflowing gas, as first tentatively reported by \cite{ruschel2021agnifs}. Considering the almost face-on orientation of NGC~4593, we most likely see the approaching side of the outflow, and its potentially red-shifted counterpart is hidden by the galaxy disc. 

\subsection{Resolved BPT diagrams: Ionisation mechanisms}
\label{bpt}

\begin{figure}
    \centering
     \includegraphics[width = 0.45\textwidth]{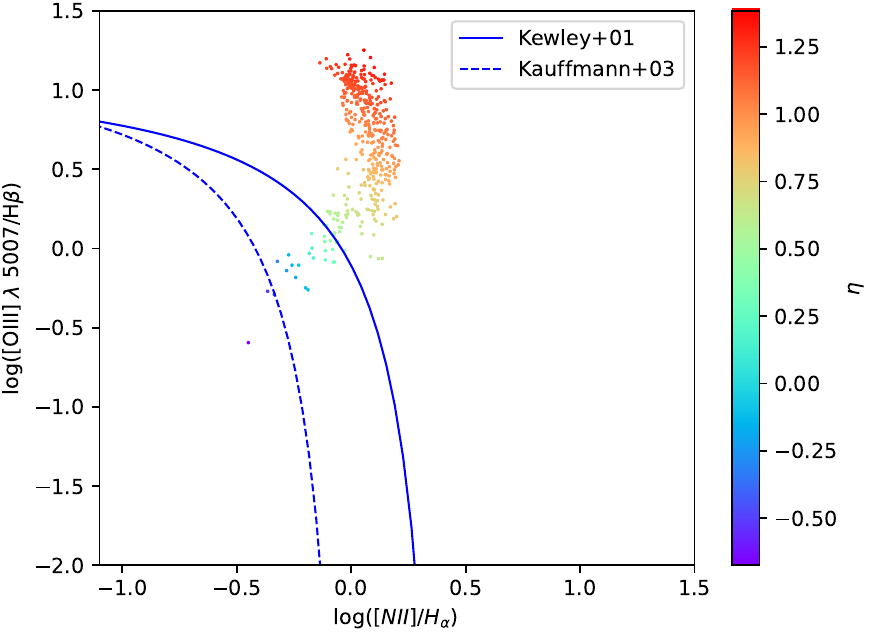}
     \includegraphics[width = 0.45\textwidth]{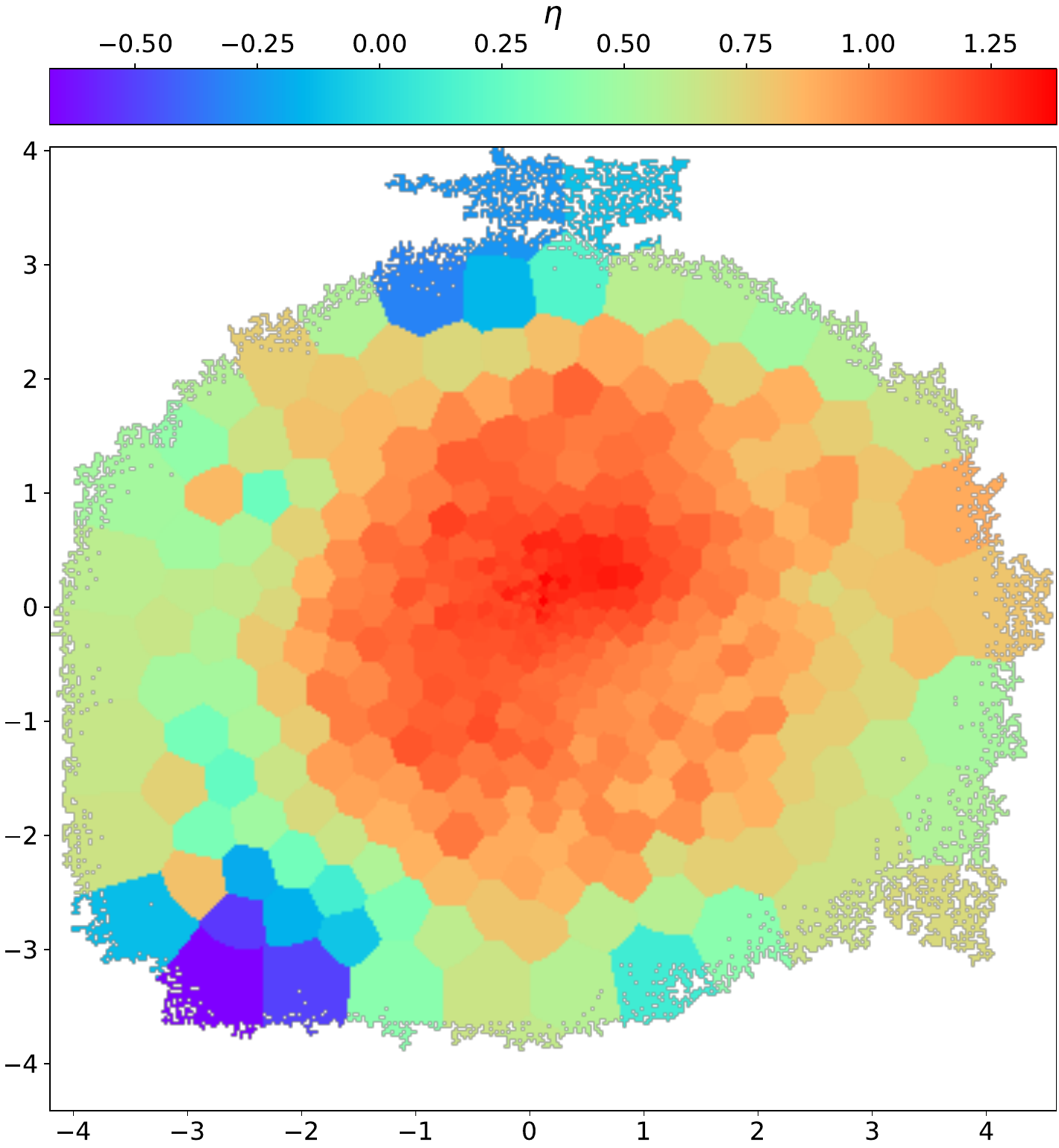}
   \caption{BPT diagram (top panel) of the regions shown in the bottom panel. The colour scale is shown, colour-coded by $\eta$, in the colour bar on the side and at the top of each map, respectively. The scale of the axes in the right panel is arcsec. North is up and east is to the left.}
  \label{Fig.5}
 \end{figure}

To study the mechanism(s) that ionise the central region of NGC~4593, we fitted the [O\,\textsc{iii}]\,$\lambda$5007 and [N\,\textsc{ii}]\,$\lambda$6583 lines with a narrow Gaussian component, and H$\alpha$ and H$\beta$ with two (narrow and broad) Gaussians. The broad-component fitted to H$\alpha$ and H$\beta$ corresponds to emission from the BLR, and the narrow Gaussian includes the disc and outflow components. We did not consider the latter as two separate kinematic components as we did in previous sections because the BLR component makes it impossible to achieve a good separation into three components for the H$\alpha$ and H$\beta$ lines. We then used the BPT  diagnostic diagram \citep{baldwin1981classification} to produce resolved ionisation diagnostic  maps. The maps used the [N\,\textsc{ii}]$\,\lambda$6583/H$\alpha$ and [O\,\textsc{iii}]$\,\lambda$5007/H$\beta$ narrow line flux ratios.
The solid curve in the left panel of Fig.~\ref{Fig.5} indicates the theoretical upper limit to pure star formation from \citet{kewley2001theoretical}, and the empirical limit to this from \cite{kauffmann2003host} is indicated with the dashed curve. We used the parameter $\eta$ defined by \cite{erroz2019muse} as an indicator of the strength of the ionisation mechanism, given by the distance of a given point to the curves. 
The region below the dashed line is considered to be ionised by star formation ($\eta < -0.5$), and the region above the solid line is ionised by the AGN ($\eta > 0.5$). The region between the two lines is considered to be composite and is ionised by a mix of the two mechanisms. 

The BPT diagram shown in the left panel of Fig. \ref{Fig.5} indicates that the bulk of the narrow-line emission measured from the MUSE data is gas that is photoionised by the AGN, with a few spaxels lying in the composite region of the diagram. The right panel of Fig.~\ref{Fig.5} shows the spatial distribution of $\eta$. It reveals higher ionisation parameter values in the central $\sim$4\arcsec(680\,pc) in diameter of the galaxy ($\eta > 1$), as well as values of 0.75<$\eta$<1 towards the west. The spaxels lying in the composite region of the diagram ($-0.5<\eta<0.5$) are found at radii larger than $\sim$2\arcsec(340\,pc).

\subsection{Physical properties of the outflowing gas}

From the [O\,\textsc{iii}]\,$\lambda$5007 emission line flux and the kinematics presented in Section \ref{emission}, we can measure various physical properties of the outflowing gas. For all the calculations in this Section, we used the measurements corresponding to the [O\,\textsc{iii}]$\,\lambda$5007 broad-components shown in Fig.~\ref{Fig.4}. To estimate the total mass of the outflow $M_{\rm out}$, we used Eq.~\ref{Eq.1} in \cite{rose2018quantifying},

\begin{equation}
    M_{\rm out} = \frac{L\Big({\rm H}\beta\Big)_{{\rm corr}}\,m_{\rm p}}{{\alpha_{{\rm H}\beta}^{\rm eff}}\,h\nu_{{\rm H}\beta}\,n_{\rm e}} ~,
 \label{Eq.1}
\end{equation}

\noindent where $L\Big({\rm H}\beta\Big)$ is the H$\beta$ luminosity corrected for extinction, $\alpha_{{\rm H}\beta}$ is the effective Case B coefficient (we adopted $\mathrm {3.03 \times 10 ^{-14}\,cm^{3}\,s^{-1}}$), which corresponds to an electron temperature $T_{\rm e} = 10^4\,{\rm K}$ \citep{osterbrock2006astrophysics}, $m_{\rm p}$ is the mass of the proton, $h\nu_{{\rm H}\beta}$ is the energy of an H$\beta$ photon, and $n_{\rm e}$ is the electron density.

\begin{figure}
    \centering
    \includegraphics[width = 0.5\textwidth]{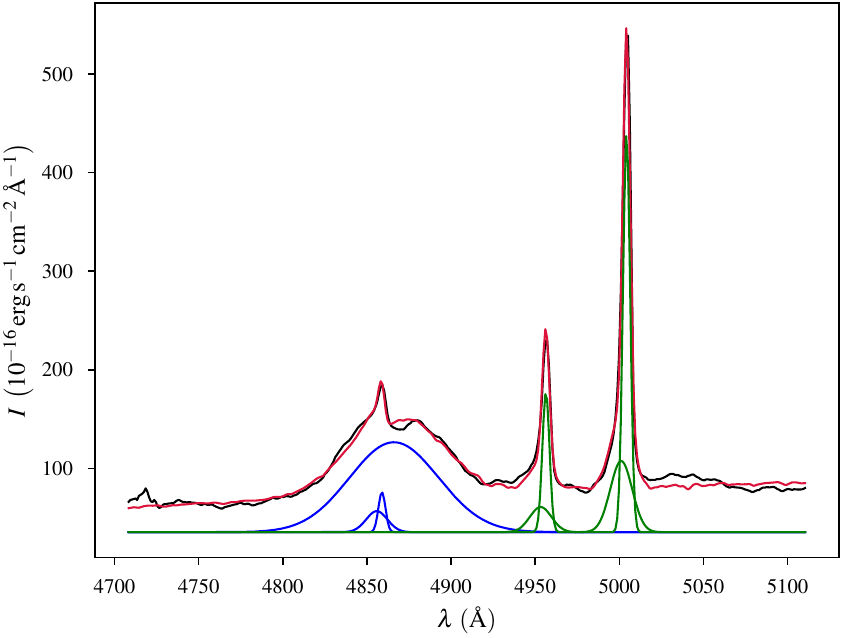}
    \caption{ H$\beta$ and [O\,\textsc{iii}]$\,\lambda$5007 emission lines and corresponding fits of the nuclear spectrum. The solid black and red lines correspond to the MUSE data and fit. The latter is the sum of three components in the case of H$\beta$: A broad-component with an ${\rm FWHM}=1600$\,km\,s$^{-1}$ associated with the BLR, a blue-shifted broad-component with an ${\rm FWHM}=400$\,km\,s$^{-1}$ representing the outflow, and a narrow-component of ${\rm FWHM}=100$\,km\,s$^{-1}$ corresponding to the gas disc. In the case of the [O\,\textsc{iii}]$\,\lambda$5007 lines, only the two latter components were fitted.}
    \label{fig:6}
     \end{figure}

To measure the H$\beta$ line flux and hence obtain its luminosity, we first tied its kinematics to those of [O\,\textsc{iii}]$\,\lambda$5007 to reduce the uncertainties. We then fitted the emission line profile leaving the intensity as the only free parameter, as shown in Fig.~\ref{fig:6}, where the blue-shifted component representing the outflow is buried below the broad-component of ${\rm FWHM}\sim$1600 km~s$^{-1}$. 
$L({\rm H}_{\beta}$) in Eq.~\ref{Eq.1} should be corrected for extinction, but from the data we could not determine the extinction using the H$\beta$ and H$\alpha$ lines because the emission from the BLR is present everywhere in the field due to point-spread function (PSF) smearing.  
As a result, the H$\beta$ luminosity that we measure, $4.7\times10^{39}$\,erg\,s$^{-1}$, and all the derived quantities are lower limits. 

\begin{figure}
    \centering
    \includegraphics[width = 0.5\textwidth]{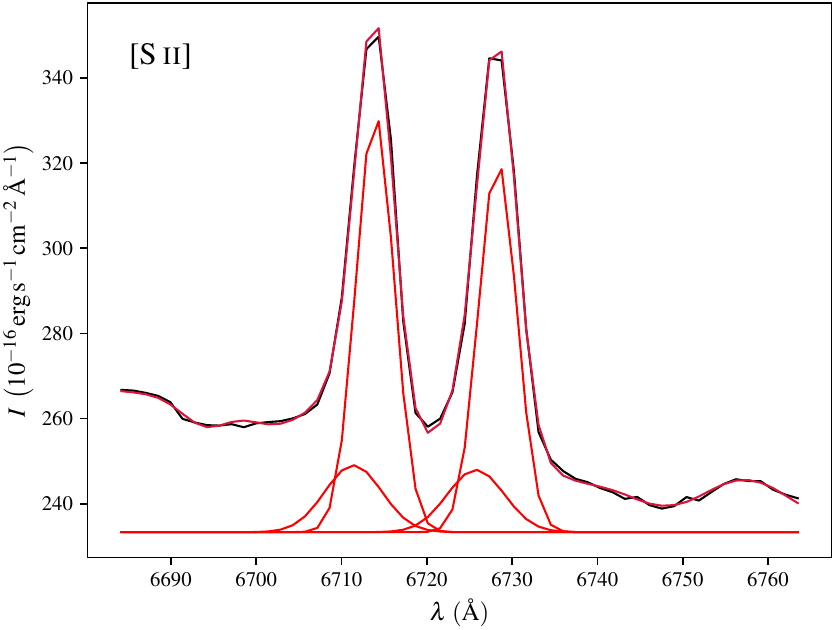}
    
    \caption{Same as in Fig. \ref{fig:6},but for the [S\,\textsc{ii}] doublet, integrated in a circle of $2\arcsec$, centred on the nucleus.}
   
    \label{fig:7}
\end{figure}

To estimate the electron density, we first attempted to create a map of the ratio [S\,\textsc{ii}]\,$\lambda$6716/[S\,\textsc{ii}]\,$\lambda$6731, but the S/N of the data was not high enough. We thus constructed an integrated [S\,\textsc{ii}] spectrum within a diameter of $2\arcsec$, centred on the galaxy nucleus (see Fig. \ref{fig:7}) as shown in \citet{sanders2015mosdef}. Using this spectrum and the flux of the broad-components that we fitted, we obtained an electron density of $n_{\rm e}= 400\,{\rm cm}^3$.
Using this electron density, we measure a total mass of the outflow of $M_{\rm out}\geq 1.589\times 10^{35}$\,kg which corresponds to $0.799\times 10^5 \,M_{\sun}$.

To obtain the mass-loss rate $\dot{M}$, we used Eq.~\ref{eq2},

\begin{equation} 
          \dot{M} = \frac{M_{\rm out} {V_{\rm out}}}{R_{\rm out}}~,
\label{eq2}
\end{equation} 

\noindent where $V_{\rm out}$ is the outflow velocity, which we estimated as the maximum velocity that we measured for the outflowing gas. In this case it is $\sim200$\,km\,s$^{-1}$, derived from the blue-shifted broad-components shown in the middle panel of Fig.~\ref{Fig.4}. The outflow radius, $R_{\rm out}$, was estimated to be 2\arcsec~(340\,pc) from the nucleus, as measured from the right panel of Fig.~\ref{Fig.4}. We adopted a model of an expanding spherical shell assumed to be accelerated at the nucleus \citep{kraemer2020mass}. Based on this we calculated that the mass-loss rate per year is $\dot{M} \geq 0.048\,M_{\odot}$. 

To estimate the kinetic power $\dot{E}_{\text{kin}} $ of the outflow, we used Eq.~4 in \citet{rose2018quantifying},

\begin{equation} 
          \dot{E}_{\rm kin} = \frac{\dot{M}_{\rm out}}{2} (V_{\rm out}^2 + 3\sigma^2),
\label{eq3}
\end{equation}

\noindent where $\sigma$ is the average velocity dispersion of the outflow measured from the right panel of Fig.~\ref{Fig.4}, of $\sim250$\,km\,s$^{-1}$. The resulting value of the kinetic power is $\dot{E}_{\text{kin}}\geq\mathrm{4.09\times 10^{39}\,{erg\,s}}^{-1}$. To estimate the coupling efficiency, we divided the outflow kinetic energy by the bolometric luminosity of the AGN, $\mathrm{5\times10^{43}\,erg\,s^{-1}}$, obtained from 2-10 keV X-ray observations of the galaxy (\citealp{vasudevan2009simultaneous}). \citet{ruschel2021agnifs} reported an AGN luminosity of $\mathrm{1\times10^{44}\,erg\,s^{-1}}$ based on more recent hard X-ray observations. Considering these two values we thus derive a coupling efficiency of $\geq 8.18\times 10^{-5}$.

 \section{Discussion}
 The superb spatial resolution of the MUSE-NFM data set used in this work is demonstrated by the MUSE image of NGC4593 collapsed in the spectral range 5300 -- 5750\,$\AA$ shown in Fig.~\ref{Fig.1}, where almost all the structures seen in an HST image are detected in the circumnuclear region. We can clearly distinguish the innermost spiral arm originating in the south, which surrounds the nucleus in anti-clockwise direction, until it reaches the nucleus through the north. Dust lanes are present in the spiral arms, and they mostly coincide with the CO(1-0) line emission detected with ALMA. The central region of the galaxy is most likely accumulating gas that will eventually fuel future episodes of nuclear activity and/or star formation.  
 
 From the stellar velocity maps derived from the MUSE data, we confirm the presence of a rapidly rotating disc. The MUSE data allowed us to study the kinematics and energetics of the ionised outflow previously reported by \citet{ruschel2021agnifs} using optical IFU data from the Gemini Multi-Object Spectrograph (GMOS). These authors tentatively reported the presence of extended [O\,\textsc{iii}]$\,\lambda$5007 emission along the photometric major axis whose emission peak is displaced about 40\,pc north-east of the nucleus. They measured the highest velocity dispersion and density values to the east and south-east, with the later reaching maximum values of $\sim$500-600 cm$^{-3}$. In the MUSE data we detect blue-shifted gas, with velocities of up to 200\,km\,s$^{-1}$, in a region with a high velocity dispersion of up to 250\,km\,s$^{-1}$ towards the east side of the galaxy. We measure an outflow radius of 2\arcsec (340\,pc) from the broad-component of the [O\,\textsc{iii}]$\,\lambda$5007 line. Considering the almost face-on orientation of NGC~4593, we most likely detect the approaching side of the outflow, and its potentially receding counterpart would be hidden by the galaxy disc. 

\begin{figure}
    \centering
    \includegraphics[width=0.5\textwidth]{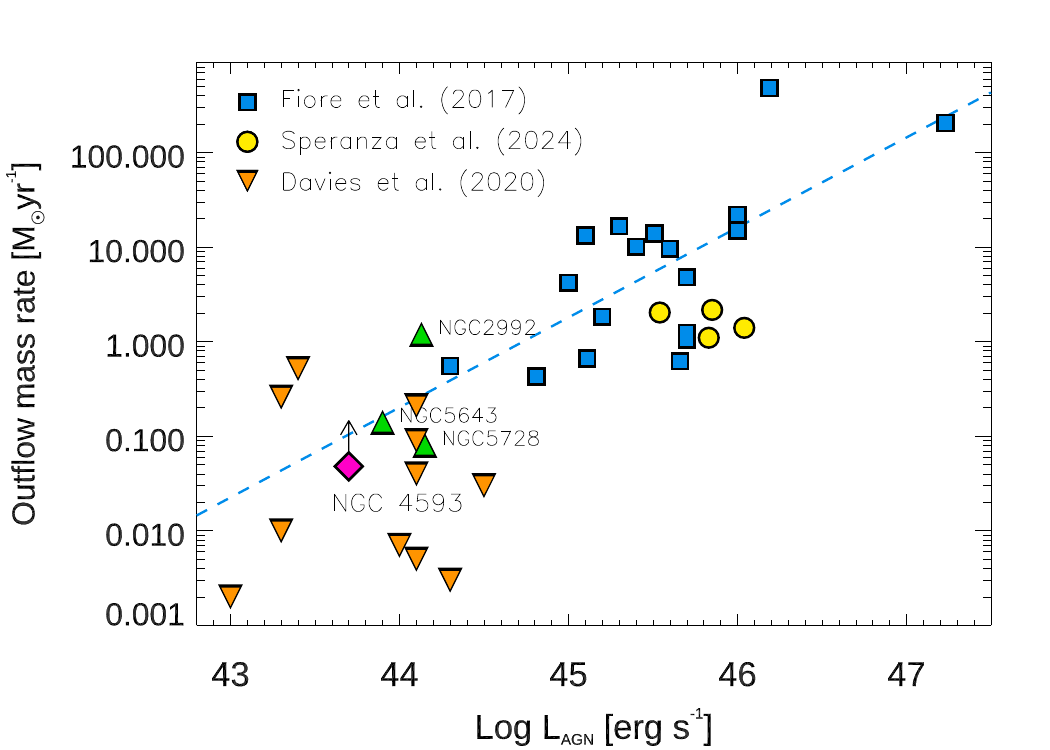}
    \caption{Ionised mass outflow rate as a function of AGN luminosity. Blue squares show the ionised outflows of ULIRGs and AGN at $z<0.5$ compiled by \citet{fiore2017agn}, and the dashed blue line is the corresponding linear fit. Yellow circles and orange triangles show the ionised outflows of the QSOFEED type 2 quasars at $z\sim0.1$ from \citet{speranza24} and of the nearby Seyfert galaxies from \citet{davies2020}. The pink diamond shows the lower limit on the outflow rate in NGC~4593 derived in this work, and the green triangles show the values reported for NGC\,2992 \citep{zanchettin2023}, for NGC\,5643 \citep{garcia2021}, and for NGC\,5728 \citep{shimizu2019}. The outflow mass rates from \citet{fiore2017agn}, \citet{zanchettin2023}, and \citet{speranza24} have been divided by three to match our outflow geometry (see Eq.~\ref{eq2}).}
        \label{fig:8}
\end{figure}

Since we cannot constrain the outflow extinction, we derived lower limits to the outflow mass, mass-loss rate, kinetic energy, and coupling efficiency. At the bolometric luminosity of NGC\,4593, a mass-loss rate of $\sim 0.1\,M_\sun\,{\rm yr}^{-1}$ is expected from the \cite{fiore2017agn} empirical relation at $z<0.5$ between outflow mass rate and AGN luminosity (see Fig.~\ref{fig:8}). The value that we estimate from the MUSE observations is $\dot{M} \geq 0.048 \,M_{\odot}\,{\rm yr}^{-1}$, which is consistent with the empirical relation. Figure~\ref{fig:8} also includes the type 2 quasars from the QSOFEED sample reported by \citet{speranza24}. These ionised outflows are below the scaling relation due to their large radii (between 3 and 12\,kpc) and high electron densities ($300-1000$\,cm$^{-3}$), measured also from the [S{\sc ii}] emission lines. The same is true for most of the Seyfert galaxies with ionised outflow measurements reported by \citet{davies2020}, which are also shown in Fig.~\ref{fig:8} and have accurate measurements of the electron density. The galaxies compiled by \cite{fiore2017agn}, were all assumed to have the same density and radius (200\,cm$^{-3}$ and 1 kpc), resulting in higher outflow mass rates. 
In the case of NGC~4593, the values of the outflow radius and the electron density are $\sim340$\,pc and 400\,cm$^{-3}$, respectively, and they also contribute to boosting the ionised mass outflow rate.

 \section{Conclusions} 

We presented a detailed analysis of the circumnuclear region of NGC~4593, a Seyfert 1.3 galaxy located at a distance of 35.35  $\pm 2.95$ Mpc. We used archival optical IFS data from the AO-assisted MUSE-NFM instrument at the VLT. Dust lanes are present in the spiral arms and most of them coincide with the CO(1-0) line emission detected in ALMA data (see Fig.~\ref{Fig.1}).
The stellar velocity and velocity dispersion maps are similar to those of the ionised gas disc (see Sect.~\ref{emission}) and show that the circumnuclear stellar population is supported by rotation with a velocity amplitude of $\pm$110\,km\,s$^{-1}$, which matches the findings by \citet{barbosa2006gemini}. 
Kinematic maps from two-component Gaussian fits of the [O\,\textsc{iii}]$\,\lambda$5007 line show blue-shifted velocities of up to $-200$\,km\,s$^{-1}$ on the east side of the galaxy, within a radius of $\sim2\arcsec$. This coincides with a region with a high velocity dispersion (up to 250\,km\,s$^{-1}$), confirming the presence of outflowing gas, as first tentatively reported by \cite{ruschel2021agnifs}. 

We measured some of the properties of the outflowing gas; a mass-loss rate of $\dot{M} \geq 0.048\,M_\odot\,{\rm yr}^{-1}$, a kinetic power of $\dot{E}_{\text{kin}}\geq\mathrm{4.09\times 10^{39}\,{erg\,s}}^{-1}$ and a coupling efficiency of of $\geq 8.18\times 10^{-5}$. The derived mass-loss rate is consistent with that predicted from the empirical relation of \cite{fiore2017agn}, $\dot{M}\sim 0.1\,M_\odot\,{\rm yr}^{-1}$ at the bolometric luminosity of NGC\,4593. As constrained from our BPT diagrams, the bulk of the line-emitting gas associated with the disc and the outflow that we detect with MUSE is photoionised by the AGN, with a few spaxels at radii larger than $\sim$2\arcsec~lying in the composite region of the diagram.

Our study has proven the extraordinary quality of MUSE data once again. Although an ionised outflow in this galaxy has been tentatively reported before by \citet{ruschel2021agnifs}, the MUSE data only allow us two-component fits at high spatial resolution, which can be used to study the kinematics of the ionised gas in this galaxy. Most studies have not been able to resolve ionised outflows for low-luminosity AGN such as NGC\,4593, but the high angular resolution of MUSE-AO, now allow us to constrain the outflow radius to be ${\rm\sim 340\,pc}$. It is expected that further high angular resolution IFU observations will provide more measurements of ionised outflows in low-luminosity AGN such as NGC\,4593. 
 \\
 \\
 {\textit{Acknowledgements}}: \footnotesize {Part of the results in this work are based on the public data released from MUSE NFM-AO under the programme ID. NO.3-0908, operated by ESO. We also used data from ALMA, as part of project 2021.1.00812.S (PI: C. Ricci)}.This research was also supported by a grant from the European Astronomical Society thanks to the generous support of the MERAC Foundation and Springer Verlag. DM thanks the Instituto de Astrof\'\i sica de Canarias (IAC) for allowing us to use their super-computers remotely to process the data used in this work Co-funded by the European Union. Views and opinions expressed are however those of the author(s) only and do not necessarily reflect those of the European Union. Neither the European Union nor the granting authority can be held responsible for them. JHK acknowledges support from the Agencia Estatal de Investigaci\'on del Ministerio de Ciencia, Innovaci\'on y Universidades (MCIU/AEI) under the grant``The structure and evolution of galaxies and their outer regions'' and the European Regional Development Fund (ERDF) with reference PID2022-136505NB-I00/10.13039/501100011033. SC acknowledges funding from the State Research Agency (AEI-MCINN) of the Spanish Ministry of Science and Innovation under the grant ``Thick discs, relics of the infancy of galaxies'' with reference PID2020-113213GA-I00. CRA acknowledges funding from the State Research Agency (AEI-MCINN) of the Spanish Ministry of Science and Innovation under the grant ``Tracking active galactic nuclei feedback from parsec to kiloparsec scales'', with reference PID2022-141105NB-I00. CRA thanks Maria Vittoria Zanchettin for helpful suggestions. 
 
\bibliographystyle{aa}
\bibliography{refs}
\end{document}